# A deterministically fabricated spectrally-tunable quantum dot based single-photon source

*Sarah Fischbach, Martin von Helversen, Marco Schmidt, Arsenty Kaganskiy, Ronny Schmidt, Andrei Schliwa, Tobias Heindel, Sven Rodt, Stephan Reitzenstein\**

Institut für Festkörperphysik, Technische Universität Berlin, Hardenbergstraße 36, 10623 Berlin, Germany

\*E-mail: stephan.reitzenstein@physik.tu-berlin.de

**Abstract**

Spectrally-tunable quantum-light sources are key elements for the realization of long-distance quantum communication. A deterministically fabricated single-photon source with a photon-extraction efficiency of $\eta = (20 \pm 2)$ % and a tuning range of $\Delta E = 2.5$ meV is presented here. The device consists of a single pre-selected quantum dot monolithically integrated into a microlens which is bonded onto a piezoelectric actuator via thermocompression goldbonding. The thin gold layer simultaneously acts as a backside mirror for the quantum dot emission, which is efficiently extracted from the device by an optimized lens structure patterned via 3D in-situ electron-beam lithography. The single-photon nature of the emission is proven by photon-autocorrelation measurements with $g^{(2)}(\tau = 0) = 0.04 \pm 0.02$. The combination of deterministic fabrication, spectral-tunability and high broadband photon-extraction efficiency makes the microlens single-photon source an interesting building block towards the realization of quantum repeaters networks.

**Main Text**

1. **Introduction**

Quantum communication protocols promise secure data transmission based on single-photon technology.[1-3] High data transmission rates in quantum key distribution can be achieved by using efficient single-photon sources, which can be realized by nanofabrication of semiconductor devices based on quantum dots (QDs).[4-9] Beyond that, implementations of long-distance quantum key distribution require Bell-state measurements in quantum repeaters[10,11] to transfer quantum states between different nodes of a communication network. Such schemes rely on entanglement swapping which can be implemented with indistinguishable pairs of entangled photons generated by remote sources.[12] Thus, to enable large scale quantum repeater networks, sources emitting at the same energy, on the order of the homogeneous linewidth of the emitters, are required in each quantum node.

Semiconductor QDs are promising candidates, as they emit single photons with indistinguishability exceeding 95 %[9, 13, 14] and polarization-entangled photon pairs with close to ideal fidelity.[15-17] However, the self-assembled Stranski-Krastanov growth mode, which is typically used to achieve high-quality QDs, leads to randomly distributed emitters with varying shape and size, resulting in an emission band with inhomogeneous broadening of typically 10-50 meV. With respect to the requirement of realizing spectrally matched single-photon sources, deterministic in-situ processing techniques[18, 19] allow one to pre-select emitters within this emission band with an accuracy of about 1 meV. Thus, an additional spectral fine-tuning knob is required to achieve spectral resonance of multiple single-photon sources within the QD's homogeneous linewidth of about 1 µeV to enable entanglement swapping in quantum repeater networks. Moreover, the precise tunability of single-photon sources is also beneficial for the coupling of single-photon emitters to other key components of advanced quantum networks, namely quantum memories, realized e.g. by atomic vapors,[20] trapped atoms[21] or solid state quantum memories.[22]

Various methods have been applied to achieve spectral control of the QD emission characteristics, often accompanied with drawbacks: Temperature tuning,[23] for instance, suffers from increased phonon-contributions finally limiting the photon indistinguishability already above 10-15 K.[14, 24, 25] Electric fields can be applied to influence the QD emission via the quantum-confined Stark effect.[26, 27] This scheme, however, requires complex doping and contact schemes which can degrade the optical performance of the device. Strain-tuning proved to be an excellent alternative, which can be implemented by integration of the emitter onto a piezoelectric material such as $Pb(Mg_{1/3}Nb_{2/3})O_3$-$PbTiO_3$ (PMN-PT).[28, 29] In addition to the spectral-tunability, strain-tuning can be used to control the exciton binding energies and the fine structure splitting of QD states, which enables the generation of polarization-entangled photon pairs.[30] In view of applications of quantum-light sources in secure quantum communication scenarios, high photon-extraction and collection efficiencies are desirable to achieve high data transmission rates. So far, only few attempts have been made to increase the efficiency of strain-tunable single-photon sources. For example, using strain-tunable nanowire antennas, an extraction efficiency of 57 % into a numerical aperture of 0.8 has been achieved.[31]

In this work, we present a bright spectrally-tunable single-photon source based on a deterministically fabricated QD microlens combined with a piezoelectric actuator by a flip-chip goldbonding technique. The applied 3D in-situ electron-beam lithography (EBL) technique has the important advantages that suitable QDs can be pre-selected by their emission energy with an accuracy of 0.25 meV before they are integrated into photonic nanostructures with an overall alignment accuracy of about 30-40 nm [32] to achieve a broadband enhancement of the photon-extraction efficiency.

## 2. Device design and fabrication

The fabrication of our device involves three processing steps: It starts with the growth of a semiconductor heterostructure by metal-organic chemical vapor deposition. Subsequently, a flip-chip thermocompression goldbonding process is applied, which results in a thin GaAs membrane including the QDs attached to the piezoelectric actuator. In a final step, single QDs are deterministically integrated into microlenses by means of in-situ EBL.

The growth process starts with an $Al_{0.97}Ga_{0.03}As$ layer with a thickness of 1 μm which is deposited on a GaAs (100) substrate, acting as an etch stop layer later on. Above this layer, 570 nm of GaAs are grown including the InGaAs QDs in a distance of 200 nm to the sample surface. For the flip-chip bonding process, 200 nm of gold are deposited onto the sample using electron-beam evaporation. Additionally, a 300 nm gold layer is evaporated on a PIN-PMN-PT $(Pb(In_{1/2}Nb_{1/2})O_3\text{-}Pb(Mg_{1/3}Nb_{2/3})O_3\text{-}PbTiO_3)$ crystal. This material is chosen as it has an increased depoling temperature of $T_C = 140\ °C$ and a higher coercive field of $E_c = 6\ \text{kV cm}^{-1}$ compared to the more commonly used PMN-PT with $T_C = 90\ °C$ and $E_c = 2.5\ \text{kV cm}^{-1}$.[33] Next, the QD sample is placed upside-down onto the piezoelectric actuator with the two gold layers facing each other (cf. **Figure 1(a)**). A pressure of 6 MPa at a temperature of approximately 600 K is applied for 4 hours to achieve a strong cohesion of the gold layers.

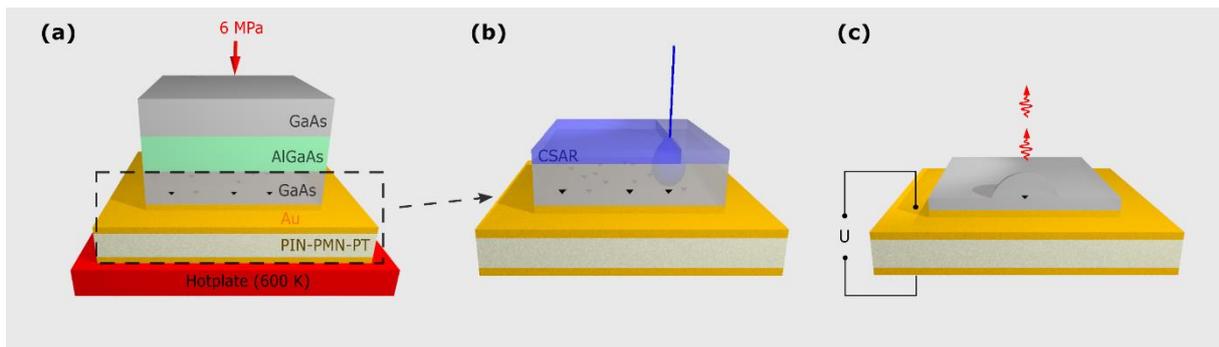

**Figure 1.** Schematic illustration of the fabrication process of a tunable QD microlens: (a) Thermocompression goldbonding of the layer structure including InGaAs QDs, followed by a wet etching step to remove the GaAs substrate and the etch stop layer. (b) Mapping process for the in-situ electron beam lithography. Suitable QDs are chosen and integrated into microlens structures. (c) The PIN-PMN-PT is contacted to transfer strain to the QD microlens for spectral-tuning of the single-photon emission.

As a next step, the upper GaAs substrate is removed by a stirred solution of hydrogen peroxide and ammonium hydroxide until the etching stops at the $Al_{0.97}Ga_{0.03}As$ layer. The latter is removed by hydrochloric acid such that a semiconductor membrane with a thickness of 570 nm remains on top of the gold layer.

To enhance the photon-extraction efficiency and to pre-select QDs with a specific emission energy, 3D in-situ EBL at 10 K is applied. This method allows us to choose QDs with a target emission energy and high emission intensity within a scanned area of the sample by their cathodoluminescence characteristics. **Figure 1(b)** illustrates the sample during the mapping process. Sample areas of 20 μm x 20 μm are scanned and suitable QDs are chosen with 0.25 meV spectral accuracy. A microlens is written into the resist on top of it, which is afterwards developed such that the structure can be transferred into the GaAs top layer by reactive-ion-enhanced plasma etching. For more details on the 3D in-situ EBL process we refer to Ref. [34]. The final device is shown in **Figure 1(c)**. The device and lens geometry was optimized beforehand using the commercially available software-package JCMsuite by the company JCMwave, which is based on a finite-element method. The optimum lens geometry is identified as a spherical segment with a height of 370 nm and a radius of 1264 nm.

### 3. Micro-photoluminescence characterization

The optical properties of the final device are investigated by means of micro-photoluminescence spectroscopy at a temperature of 10 K with a spectral accuracy of 27 μeV. **Figure 2(a)** shows a spectrum of a QD microlens device at saturation of the excitonic lines. Excitation-power- and polarization dependent measurements are used for the assignment of the emission lines to respective quantum dot states. The most intense line at $E_{X^-} = 1.3520$ eV is identified as a charged excitonic transition (X⁻), the transition at $E_X = 1.3536$ eV as the neutral excitonic transition (X) due to its polarization splitting of $\Delta E_{FSS} = 7$ μeV, while a charged biexcitonic line is observed at $E_{XX+/-} = 1.3490$ eV. To evaluate the photon-extraction efficiency $\eta$ of the microlens device, we use a pulsed diode laser ($f = 80$ MHz) to excite the QD state X⁻ at saturation and detect the emitted photons using a calibrated experimental setup (cf. Experimental Section). At zero bias voltage applied to the piezo element we observe $\eta(X^-) = (17 \pm 2)$ % for the charged excitonic transition with a linewidth of 46 µeV (FWHM).

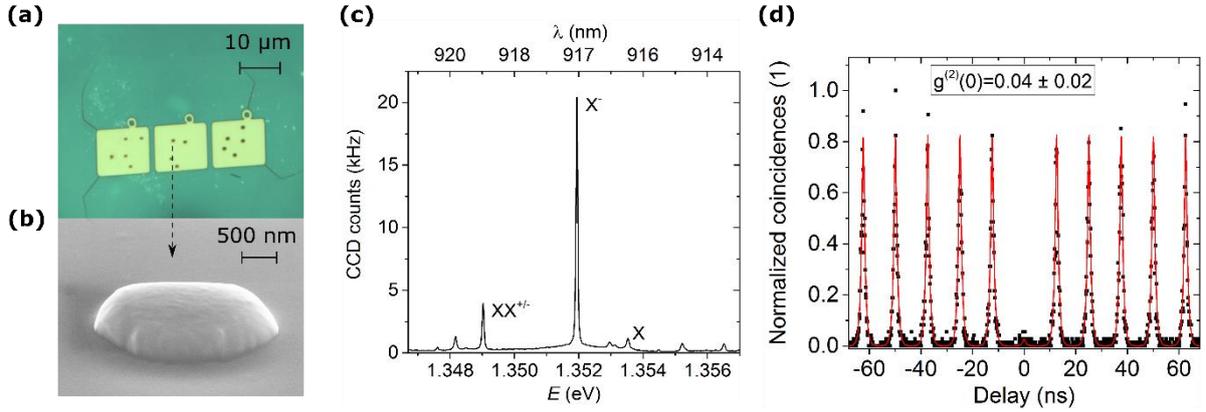

**Figure 2.** (a) Microscope image of maps taken during in-situ EBL with QD microlenses. (b) Scanning electron microscope image of a microlens. (c) Micro-photoluminescence spectrum of a QD microlens at T = 10 K. (d) Photon-autocorrelation measurements indicating a single-photon emission with $g^{(2)}(\tau = 0) = 0.04 \pm 0.02$.

Next, we verify the single-photon emission of our spectrally-tunable microlens device under pulsed wetting-layer excitation at $\lambda = 897$ nm. The photon-autocorrelation measurement at saturation of the X⁻ line in **Figure 2(d)** shows a clear antibunching at $\tau = 0$. To quantitatively evaluate the suppression of two-photon emission events, the experimental data was fitted with a sequence of equidistant two-sided exponential functions convoluted with a Gaussian of 300 ps width, accounting for the timing resolution of the Hanbury-Brown and Twiss setup. The ratio of the peak amplitudes at zero-time delay and at finite time delays reveals the second-order photon-autocorrelation $g^{(2)}(\tau = 0) = 0.04 \pm 0.02$. These results confirm that our rather complex multi-step device processing enables the realization of bright single-photon sources with a high suppression of multi-photon emission events.

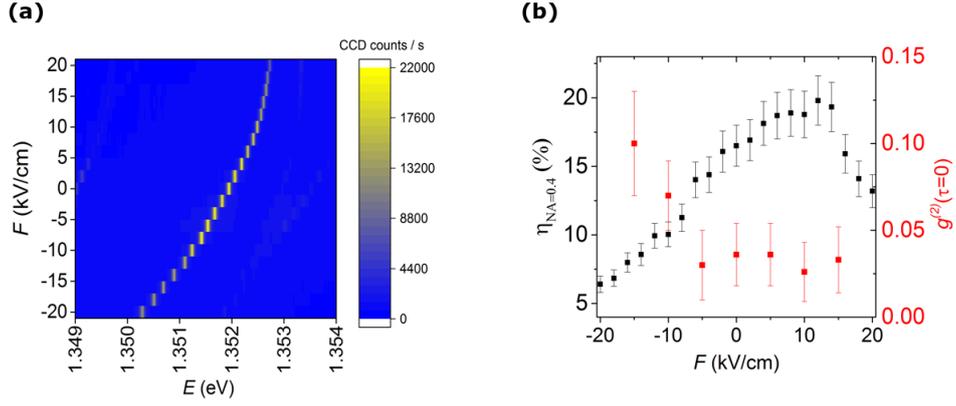

**Figure 3.** (a) Energy tuning of the X⁻ emission line by application of an electric field $F$ to the piezoelectric actuator. (b) Extraction efficiency (black, left axis) and equal-time second-order photon autocorrelation ($g^{(2)}(\tau = 0)$) results (red, right axis) for the full tuning range.

### 4. Tunability of the single-photon emission

To demonstrate the spectral tunability of QD emission, an electric field $F$ of -20 to +20 kV cm⁻¹ is applied to the PIN-PMN-PT material, corresponding to a voltage of -600 to +600 V. A positive (negative) voltage corresponds to an in-plane compression (extension) of the piezoelectric crystal transferred to the semiconductor material and the QD layer. Using the full tuning range results in a shift of the X⁻ emission by $\Delta E = 2.5$ meV as shown in **Figure 3(a)**.

To further analyze the effects of the external strain we compare our measurements to results obtained by theoretical modeling of the microlens device. The strain is modulated by adjusting the lattice constant $a_0$ of the lowest GaAs layer above the gold mirror to $\tilde{a} = a_0 - c \cdot a_0$, and the strain distribution inside the full GaAs device is calculated using continuum elasticity. One has to distinguish between the permanent strain caused by the inherent lattice mismatch between the GaAs substrate and the InGaAs QD, and the effects of the external strain caused by the piezo-tuning. Moreover, a hydrostatic strain component can be identified in contrast to a biaxial strain component, which describes a contrary impact in the lateral directions than in the vertical. **Figure 4** shows the calculation results for the permanent strain without external influence ((a1) and (b1)) as well as the additional strain effects induced by an applied external compressive as well as tensile strain ((a2) and (b2)). The distribution across the lens structure is almost uniform, only a slight relaxation effect is visible for the hydrostatic strain component as compared to the planar area around the lens.

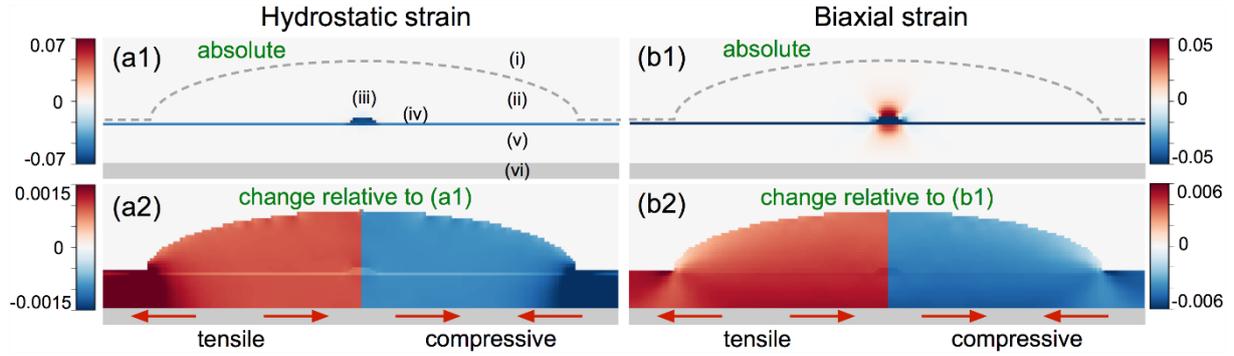

**Figure 4.** Calculated hydrostatic (a1/a2) and biaxial (b1/b2) strain distribution in a QD microlens. (a1/b1) refer to the situation in absence of external strain, while (a2) and (b2) show the additional effects by external tensile (left) and compressive strain (right). The domain is divided into (i) air, (ii) lens, (iii) QD, (iv) wetting layer, (v) spacer layer, and (vi) the piezoelectric actuator.

Additional strain may affect the energies of the localized electronic states via (i) deformation potentials, thus, changing the local band position, (ii) the alteration of the quantization energies, and (iii) the change in electron-hole Coulomb interaction. Careful analysis using eight-band k·p theory together with the configuration interaction method,[35] however, revealed that effect (i) constitutes the governing contribution, whereas (ii) and (iii) are only minor contributions, which are neglected in the following discussion. The achieved tuning of $\Delta E = 2.5$ meV corresponds to a change in the lattice constant of $c = \pm 1.2 \cdot 10^{-3}$ for compressive (+) and tensile (-) strain. At the position of the QD the resulting sum of the relative hydrostatic and biaxial strain components in all three directions are calculated separately to $\Delta \epsilon_{hy}(c) = \pm 8.1 \cdot 10^{-4}$ and $\Delta \epsilon_{biax}(c) = \pm 4.65 \cdot 10^{-3}$, where the hydrostatic strain is responsible for band-shifts and the biaxial strain for the heavy-hole light-hole splitting.[36] The sum of both effects is driving the change in the luminescence energy. Combined with the deformation potentials in $In_{0.7}Ga_{0.3}As$,

$a_g = -6725.9$ meV for the hydrostatic strain and $b_v = -1897.2$ meV for the biaxial strain, the energy shift can be calculated as

$$\Delta E(c) = a_g\big(\Delta\epsilon_{hy}(c)\big) - \frac{1}{2}b_v\big(\Delta\epsilon_{biax}(c)\big) = \pm 1.25 \text{ meV}.$$

Using the piezoelectric coefficient $d_{31} \approx 1500$ pC N$^{-1}$ as published by the manufacturer (CTS Corporation), we can compare the theoretically evaluated strain with the experimentally applied value. The maximum strain that is induced in one lateral direction during the measurement can be estimated to

$$\epsilon^{exp} = d_{31} \cdot F_{max} = 1500 \text{ pC N}^{-1} \cdot 20 \text{ kV cm}^{-1} = 3 \cdot 10^{-3},$$

compared to the theoretical value of $c = 1.2 \cdot 10^{-3}$. Matching the calculation results with the achieved tuning, it can be estimated that a fraction of $\frac{c}{\epsilon^{exp}} = 40\ \%$ of the strain effect at the piezoelectric crystal is transferred to the GaAs sample.

Besides the tunability of the emission energy, **Figure 3(a)** also reveals a change in the emission intensity with the applied electric field, which we further investigated by measuring the extraction efficiencies in pulsed excitation. As can be found in **Figure 3(b),** the highest efficiency is achieved at an applied field of $F_{max} = 12$ kV cm$^{-1}$ with $\eta(X^-, F_{max}) = (20 \pm 2)\ \%$. The efficiency decreases down to $\eta(X^-, F_{min}) = (6 \pm 1)\ \%$ at the lowest field value $F_{min} = -20$ kV cm$^{-1}$. Additionally, we investigated the second-order photon autocorrelation function for different detunings. The suppression of multi-photon emission events $g^{(2)}(\tau = 0)$ remains constant and below 0.05 over a wide tuning range and increases at high negative electric fields to $g^{(2)}(0) = 0.10 \pm 0.03$ at $F = -15$ kV cm$^{-1}$. Here, the increased $g^{(2)}(0)$ value at negative fields is attributed to enhanced uncorrelated background emission due to smaller signal to noise ratio which is not considered in the evaluation of the $g^{(2)}(0)$ data. The strain influence on the extraction efficiency could be connected to electric fields caused by charge states on the surface of the microlens. They create a field distribution around the QD which depends on the external strain. Previous studies showed that the processing of microstructures by in-situ EBL gives a lateral positioning accuracy of 34 nm.[32] Such a deviation from the center could be sufficient for the QD to be influenced by the mentioned strain-induced electric field distribution, leading to a slight separation of the electron and hole wavefunction, which in return can reduce the emission rate as we observe in the experiment. A more detailed description would require a detailed knowledge of the QD position in the microlens which is beyond the scope of the present work.

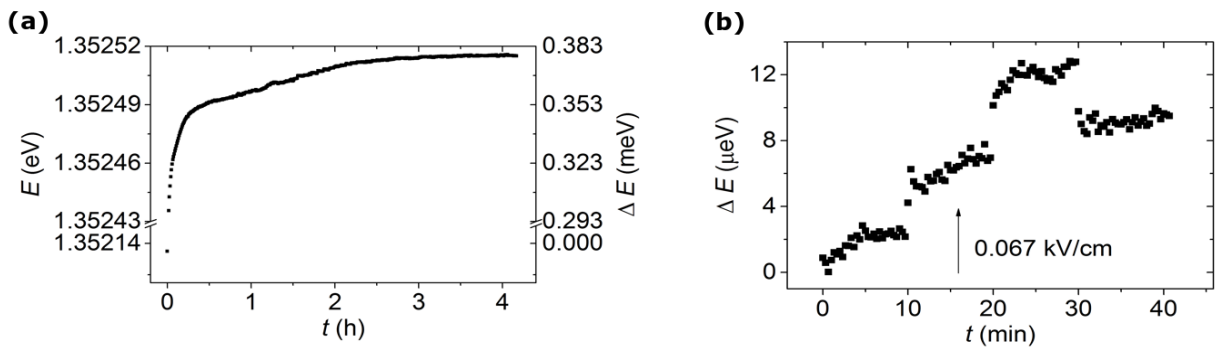

**Figure 5.** (a) Time series of the emission energy after application of the field strength $F = 12$ kV cm$^{-1}$ to the piezoelectric actuator. (b) Time-dependent change in the emission energy with small variations in the field applied to the piezo after a first stabilization as shown in (a). Two changes by $\Delta F = +\,0.067$ kV cm$^{-1}$ and one change by $\Delta F = -\,0.067$ kV cm$^{-1}$ can be observed.

In our case, the strain-dependent change of the emission characteristics does not harm the usability of the tunable source, as the available tuning range is much larger than the required one: Due to the spectral accuracy ($\pm\,0.25$ meV) of the deterministic processing technique described above, a strain-tuning range of $\Delta E = 0.5$ meV is sufficient to achieve any desired emission energy within the inhomogeneously broadened emission band of the QD emitters on the sample. Thus, for future applications we can limit the tuning to field values where high extraction efficiencies and a low probability for multiphoton emission events are provided.

A critical aspect when using piezoelectric actuators is the creep behavior of the induced strain in time, which was previously observed to follow a logarithmic function.[37] A time-dependent measurement of the emission energy of the strain-tunable QD-microlens device is presented in **Figure 5(a)**, where a field of 12 kV cm$^{-1}$ has been applied at time $t = 0$. In the first hour of the measurement series, a logarithmic time-dependency occurs. Subsequently we observe a linear drift for about 2 hours because of the typical creeping behavior of piezo-materials, before the emission finally stabilizes. Thus, for applications requiring large tuning ranges, a stabilization time of approximately 3 hours needs to be considered before operation. Once the piezo has stabilized, however, small adjustment steps can be implemented on a much shorter timescale, as indicated in **Figure 5(b)**. Changes in the tuning field by $\Delta F = 0.067$ kV cm$^{-1}$ allow for changes of $\Delta E = (4 \pm 1)$ μeV, which stay constant after less than ten minutes. Importantly, $\Delta E$ is on the order of the homogeneous linewidth of InGaAs QDs, which will facilitate quantum interference experiments and Bell-state measurements between remote emitters. To further improve the long-term stability for such experiments and future applications, it is desirable to implement an active feedback loop with PID control.

5. **Conclusion**

In conclusion, we presented a spectrally-tunable single-photon source with a maximum efficiency of $\eta = (20 \pm 2)$ % and a tuning range of $\Delta E = 2.5$ meV. The emission energy of our device is pre-selected by using in-situ EBL applied to a planar sample bonded onto a piezoelectric actuator via flip-chip thermocompression goldbonding. The tuning can serve to adjust the emission to meet the exact transition energy required e.g. for entanglement distribution in multi-node quantum networks or for the interfacing of QD based single-photon sources with atomic quantum memories. In future steps, tunable QD microlenses can be equipped with electrical contacts[38] and their emission can be collected with optical fibers,[39] to create a tunable plug-and-play single-photon source as a versatile device for quantum communication applications.

**Experimental Section**

*In-situ electron beam lithography:*

With the in-situ electron beam lithography step, QDs are chosen by their cathodoluminescence signal and integrated into microlens structures. The samples are prepared by spin-coating with the electron-beam resist AR-P 6200 (CSAR 62) and mounted onto the cold finger of a He-flow cryostat of a customized scanning electron microscope for low-temperature operation at 10 K. The reaction of the resist during development depends on the applied electron dose during exposure.[40] This resist has a positive-tone regime at low electron doses, which is used to scan small areas of the sample. The luminescence signal is focused onto a spectrometer and detected with a Si charge-coupled device camera. Based on that data, QDs are chosen and microstructures are written into the resist above them with a higher electron dose. Above a certain threshold value, the resist changes to the negative tone regime, such that the structures remain after development. The transition range to that regime is used to create quasi-3D designs. Finally, dry etching is performed by inductively coupled-plasma reactive-ion etching.

*Micro-photoluminescence measurements:*

The sample is mounted in a helium-flow cryostat and cooled down to 10 K. It is optically excited using a Titan-Sapphire laser that can be operated in quasi-continuous wave (cw) or pulsed ($f = 80$ MHz) mode. The photoluminescence is collected using a microscope objective with an NA of 0.4 and spectrally dispersed by a grating monochromator, before it is detected using a Si charge-coupled device camera. The setup is also equipped with a fiber-coupled

Hanbury-Brown and Twiss setup using single-photon counting modules based on Si avalanche photo diodes. To evaluate the extraction efficiency into the first lens of our experimental setup, the transmission of the complete setup was measured to be $\eta_{Setup} = (1.1 \pm 0.1)$ % following the procedure described in Ref. [7]. Using a laser with repetition rate $f$ a detected count-rate $n_{QD}$ corresponds to a photon-extraction efficiency of $\eta_{QD} = \frac{n_{QD}}{\eta_{Setup}*f}$ .

**Acknowledgements**

We acknowledge funding from the German Federal Ministry of Education and Research (BMBF) through the VIP-project QSOURCE (Grant No. 03V0630), the German Research Foundation via the projects SFB787 and Re2974/8-1, and from the project EMPIR 14IND05 MIQC2 (the EMPIR initiative is co-funded by the European Union's Horizon 2020 research and innovation programme and the EMPIR Participating States).